\documentclass[intlimits,twoside,a4paper]{article}

\usepackage[cp1251]{inputenc}
\usepackage[eqsecnum]{cmpj3}

\usepackage{bm}

\def\be{\begin{equation}}
\def\ee{\end{equation}}
\def\bea{\begin{eqnarray}}
\def\eea{\end{eqnarray}}

\newcommand{\rf}{\rm ref}

\newcommand{\kB}{k_{\rm B}}

\newcommand{\PM}{\rm PM}

\newcommand{\dd}{{\rm d}}

\newcommand{\HB}{{\rm HB}}

\issue{2021}{24}{3}{33501}
\doinumber{10.5488/CMP.24.33501}

\title[Thermodynamic perturbation theory and equation of state developments]%
{Thermodynamic perturbation theory and equation of state developments}

\author[I. Nezbeda]{I. Nezbeda\orcid{0000-0003-1882-0587}}

\address{Institute of Chemical Process Fundamentals, Czech Acad. Sci., 
16502 Prague 6, Czech Republic,\\
Faculty of Science, J. E. Purkinje University,
    400 96 \'{U}st\'{\i} nad Labem, Czech Republic}

\date{Received June 06, 2021, in final form July 15, 2021}
\begin{document}

\maketitle

\begin{abstract}
An alternative way of utilizing the thermodynamic perturbation
theory of Wertheim for the development of equations of state for
associating fluid models is presented and detailed for water. The
approach makes use of general features of the parameter of non-saturation to
avoid the necessary solution of an algebraic equation and unbinds the
results from a tight dependence on details of the simple reference fluid
model used in the perturbation theory.


\end{abstract}
%


\section{Introduction}
\label{Intro}


Because of its indispensable role, both for life and in industry, water has been in the focus of research for centuries. It exhibits a number of anomalies which are, more or less, well understood but for applications it is also necessary to express its properties by means of closed analytic expressions.
Such expressions may be obtained by a parametrization of available experimental data as, e.g., the IAPWS-90 equation~\cite{IAPWS} but the applicability of such equations is limited only to the range of the data used in the parametrization.
The other possibility is the use of statistical mechanics which provides a link between the atomistic view of matter and the observed macroscopic behavior. Regardless of details, equations  of state (EoS), both strictly theoretical or semiempirical, possess the form of a perturbed equation,

\begin{equation}
z \equiv \frac{\beta P}{\rho} = z_{\rf} + \Delta z,
\label{zpert}
\end{equation}
where $z$ is the compressibility factor, $P$ is the pressure, $\beta$ is the inverse temperature, $\beta=1/\kB T$, where $\kB$ is the Boltzmann's constant and $T$ is the temperature, $\rho$ is the number density, $\rho=N/V$, $z_{\rf}$ is the compressibility factor of a certain reference system, and $\Delta z$ is a correction term.
The above form of the EoS results from a decomposition of the total intermolecular interaction model into a reference part, $u_{\rf}$, and a perturbation part, $\Delta u$,

\begin{equation}
u(1,2)= u_{\rf}(1,2) + \Delta u(1,2),
\label{usplit}
\end{equation}
where (1,2) stands for a set of generalized coordinates.

Within the spirit of perturbation theories of liquids, the reference
system has to reproduce the structure of the original liquid, and its
properties should be available in an analytic form. As regards the
associating liquids, simple models exhibiting hydrogen bonding
(H-bonding), usually referred to as primitive models (PM), are employed to
meet this requirement. Such simple models emerged in the end of the
1980's~\cite{DAmodel,Bolmodel,SNmodel,KNmodel} and, nearly
simultaneously, theories of these models were developed, a theory of
Dahl and Andersen~\cite{DAtheory} and the thermodynamic perturbation
theory (TPT) of Wertheim~\cite{TPT1,TPT2}. These two ingredients
together, models and theory, opened up the way to the development of
molecular-based equations following either (i)~truly molecular approach
which starts with a realistic force field and then, by means of a
sequence of well defined approximations, a simple (primitive) model
amenable to a theoretical treatment is constructed or (ii)~a van der
Waals (vdW)-like approach which does not refer to any specific
interaction model but only makes use of the previously obtained knowledge
and constructs a reference intuitively/empirically. Both approaches use the TPT to obtain the properties of the reference model. Despite
the difference, both approaches may end up, formally, with the same EoS.

Despite all the effort invested into the development of an {\it
accurate} and {\it reliable} EoS for water, no satisfactory results have
been achieved so far. A theoretical route seems to have culminated with
an accurate description of a short-range reference, i.e., primitive
model descending from the TIP4P force field~\cite{Jirsak1,Jirsak2}; the
equation reproduces the main anomalies exhibited by water but the attempts
to use it as a reference for the development of a truly molecular-based
EoS for water were not fully satisfactory~\cite{JJ1,JJ2,JJ3}. The
vdW-like approach is represented by statistical association fluid theory (SAFT) equations. As Llovel and Vega
stated in their recent review~\cite{Vegareview}, there were 47 SAFT
equations for water available, all of them based on a sort of a
primitive model. Nonetheless, none of them meets the demands for widespread and reliable applications which only witnesses to their empirical nature.

Application of the TPT is not straightforward. The property which links
the model with the theory lies in certain integrals $I$, which should to be
evaluated for the model used. They involve correlation functions of the
reference model and the Mayer function of the H-bonding interaction (for
details see the next section). To evaluate these integrals analytically,
certain approximations should be used. Once the integrals are known,
then it is necessary to evaluate the key function of the theory --- a
non-saturation parameter $\gamma$. This evaluation lies in solving an algebraic
equation whose order depends on the order of the used TPT.

Attempting to improve the performance of the EoS developed by the
intuitive vdW-like approach, there are no other ways than (i)~to
modify/change the reference model, or (ii)~to improve the performance of
the TPT, and/or (iii)~to play games with the correction terms in the same
way as in the development of hundreds of cubic equations. As for the two
former ways, it is evident that all the computations should be always
performed again from the scratch and there is a question whether
this effort is worth doing because the imperfectness of the reference
model should be also reflected in the final results anyway. The
theoretical approach offers another possibility. Results based on the
actual underlying physics should be {\it qualitatively} the same
regardless of their parameter values. It means, the resulting functions
should exhibit the same course/functional dependence. Consequently,
focussing on general features of the derived properties as, e.g., the
parameter of non-saturation, we may skip over the details of the
reference model and thus release the direct dependence of the EoS on the
reference model. Furthermore, this approach will also introduce
additional adjustable parameters, in addition to those defining the
reference model, which should increase the chance of the final equation
to perform better.

The purpose of this paper is to examine the properties of the derived functions and to find their general course, if possible at all. At a feasibility level, the suggested route towards an EoS is examined by considering a model descending from the TIP4P force field, the model with probably the most appropriate geometry of the water molecule. In the next section we provide the basic definitions and then in the subsequent section the results. The final section contains a summary where a future potential development is also briefly outlined.

\section{Basic methodological details}

The thermodynamic perturbation theory is not a general theory since it requires a corresponding form of intermolecular interactions,

\begin{equation}
u_{\PM}(1,2) = u_0(1,2) + \sum_{j\in\Gamma}\sum_{k\in\Gamma}
         u_{\HB,jk}(1,2),
\label{upot}
\end{equation}
where $u_0$ is a reference system interaction potential, usually the hard sphere model, and the second term is a site-site interaction giving rise to H-bonding represented by a short-range attractive site-site interactions $u_{\HB}$ between sites $i$ and $k$ from set $\Gamma$. For water, typically $|\Gamma|=4$ if all sites (i.e., both those bearing a charge and uncharged ones) are counted but this value is not strictly defined and may also be both 2 and~3 in other models~\cite{SAFT-Jackson}.

The basic quantity of the theory are fundamental integrals involving the Mayer functions of the H-bonding interaction between sites $i$ and $j$,

\begin{equation}
f_{ij}(r_{ij})=\exp{[-\beta u_{\HB,ij}(r_{ij})]} - 1,
\label{Mayer}
\end{equation}
and correlation functions, $g$, of the reference system.
Denoting by (+) the hydrogen-like sites and by ($-$) the oxygen-like sites, then the 1st order theory requires only integral $I_{+-}$ defined as

\begin{equation}
I_{+-} = \int f_{+-}(r^{12}_{+-})g_0(1,2) \dd(1)\dd(2),
\label{Iij}
\end{equation}
where $g$ is the pair correlation function.
The actual form of $I$ in the second order of the TPT depends on additional approximations. In the 2nd order TPT as developed by Slovak et al.~\cite{TPT-Slovak}, integral $I$ contains the triplet correlation function, $g_{ijk}$:

\begin{equation}
I_{+-+} = \frac{1}{2}\int f_{+-}(r^{12}_{+-})f_{+-}(r^{23}_{+-})g_0(1,2,3)\dd(2)\dd(3).
\label{Iijk}
\end{equation}

The basic quantity of the theory is a parameter of non-saturation, $\gamma$. It is obtained by solving an algebraic equation with coefficients given by integrals $I$ and whose order depends on the level of approximations. For single bonding models it is a quadratic equation; for double bonding models within the second order it is, in general, an equation of the sixth order but it can be reduced, after introducing an additional approximation, to a cubic equation; for details see, e.g.,~\cite{TPT-Slovak} or a pedagogical introduction to the TPT by Zmpitas and Gross~\cite{Gross}. An additional quantity appearing in the final expressions, but which is a function of $\gamma$, is a parameter $x_0$ (in notation of Slovak et al.~\cite{TPT-Slovak}). In a general case of molecules containing $m$ hydrogen-like $(+)$-sites satisfying the condition of single bonding, and one complementary $(-)$-oxygen-like site which may form up to two bonds, the Helmholtz free energy, $A$, of the reference fluid assumes, within the TPT, the form

\begin{equation}
\beta\Delta A_{\rm PM}/N =
 \beta\Delta A_0/N + m(1-\nu) + \ln x_0   \,\,.
 \label{Afin}
\end{equation}
Integrals $I_{+-}$ and $I_{+-+}$ were evaluated by simulations and consequently parametrized so that all information necessary for the analytic evaluation of parameters $\gamma$ and $x_0$, and hence the Helmholtz free energy and all other thermodynamic properties of the primitive model, is available.

\section{Results and discussion}

To examine the above outlined idea on the application of the TPT, we consider the primitive model descending from the TIP4P force field as developed by Vlcek and Nezbeda~\cite{PM-Vlcek} along with analytic parametrizations of integrals $I$.
We have evaluated parameters $\gamma$ and $x_0$ along 9 isotherms within the range (300--700)~K for packing fractions up to a typical water value $\eta=0.35$.

\begin{figure}[h!]
\centerline{\includegraphics[scale=0.7]{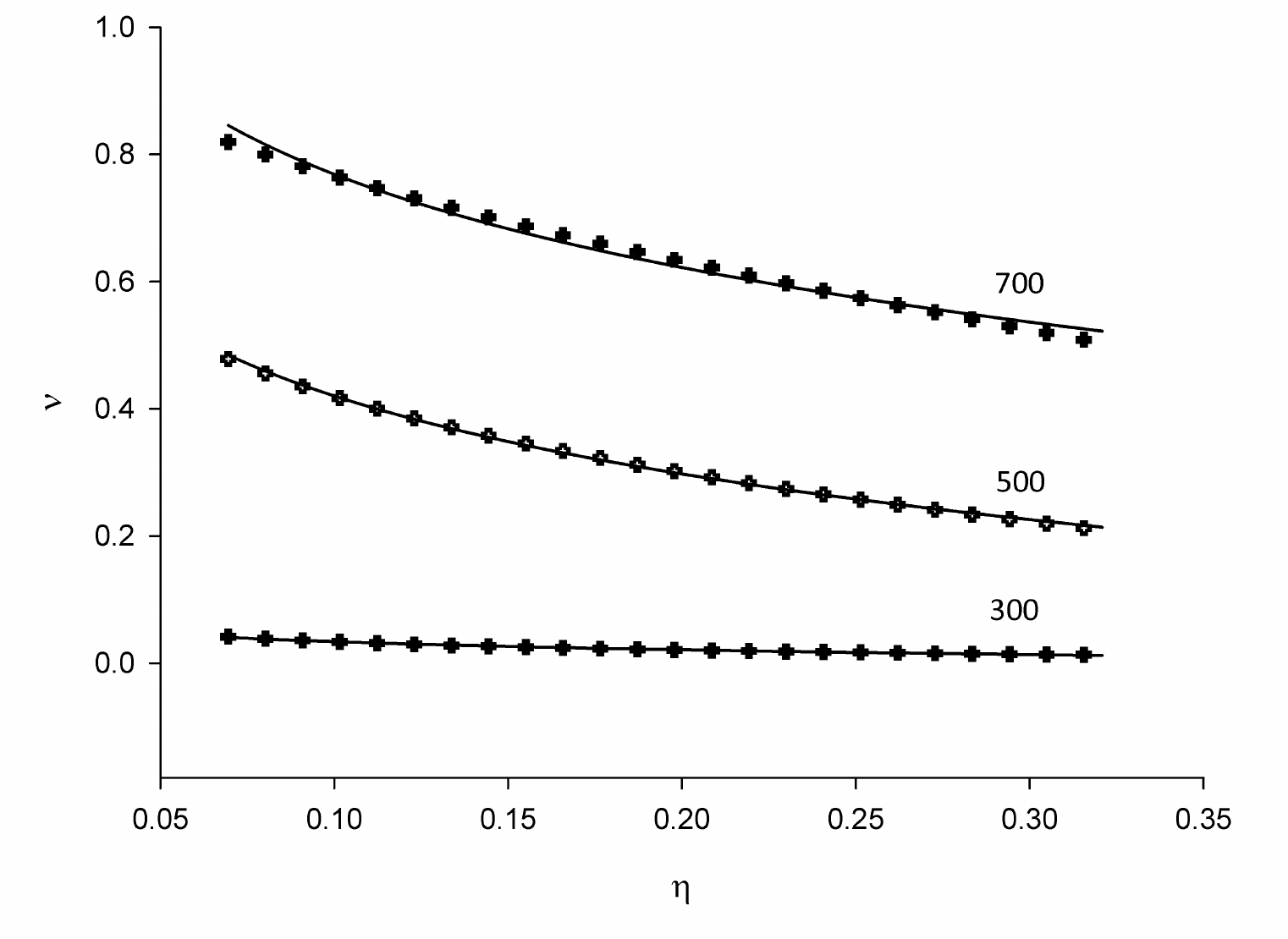}}
\caption{Parameter $\nu$ in dependence on $\eta$ as obtained from the TPT (circles) and its logarithmic parametrization (solid line). Number at curves denote temperature.}
\label{fig:nu}
\end{figure}

Parameter $\gamma$ obtained from numerical computations along with its parametrization is shown for three selected temperatures in figure~\ref{fig:nu} in dependence on density for three temperatures.

\begin{figure}[h!]
\centerline{\includegraphics[scale=0.7]{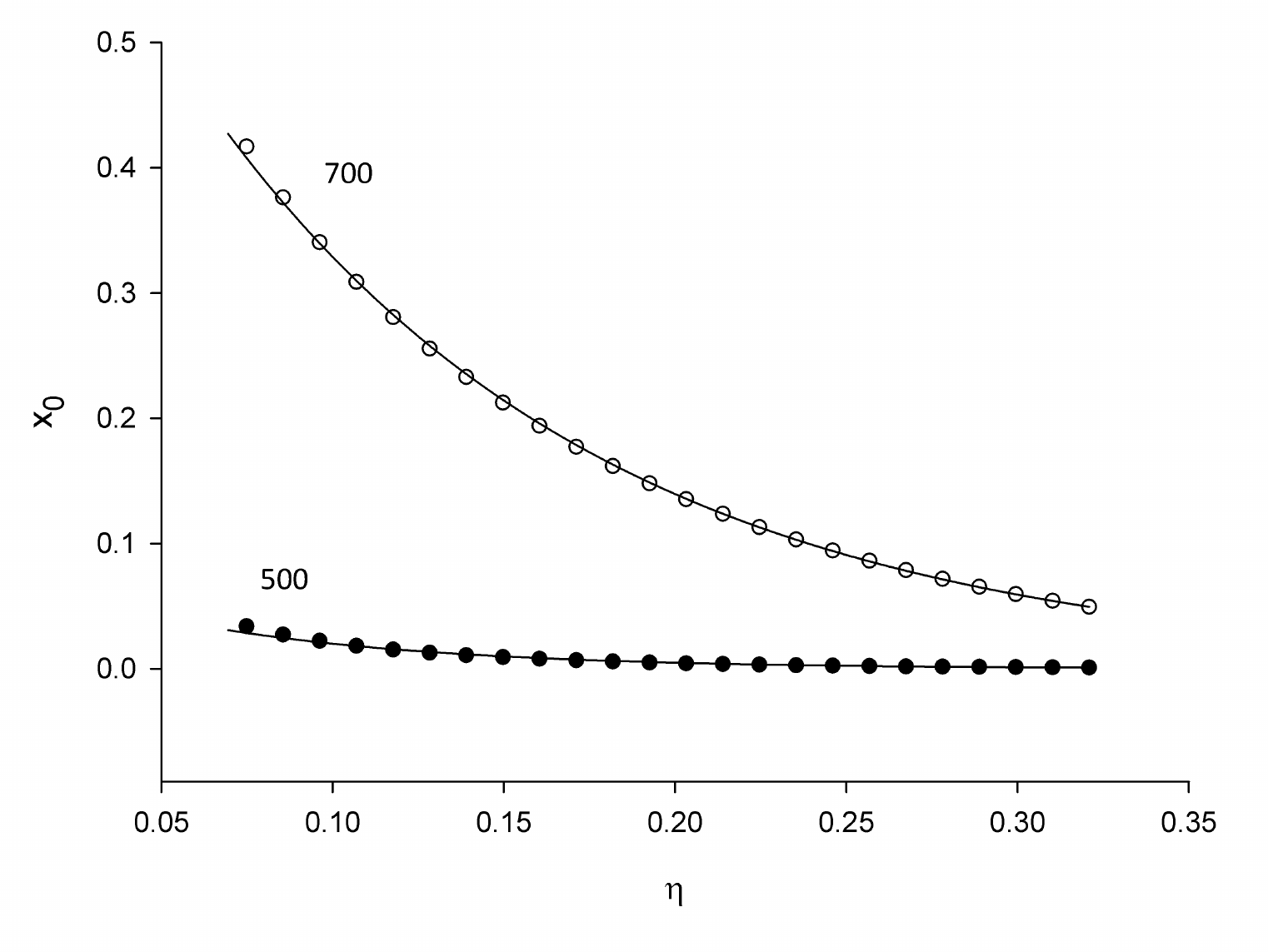}}
\caption{The same as figure~\ref{fig:nu} for $x_0$.}
\label{fig:x0}
\end{figure}

As it is seen, the logarithmic fit,

\begin{equation}
\nu(T,\rho) = s_1(T) + s_2(T)\ln(\eta)  \,\,,
\label{nu}
\end{equation}
provides an excellent fit to the numerical data and the same also applies to the exponential parametrization of parameter $x_0$,

\begin{equation}
x_0 = X_1(T) \exp[X_2(T)\eta],
\end{equation}
shown in figure~\ref{fig:x0}. The value of $x_0$ for $T=300$ is around zero and would not be thus distinguishable from a zero straight line within the scale of the graph. In fact, the exponential dependence on $T$ should not be so surprising because the underlying integrals $I$ themselves can be factorized into $T$- and $\eta$-dependent terms.

\begin{figure}[h!]
\centerline{\includegraphics[scale=0.7]{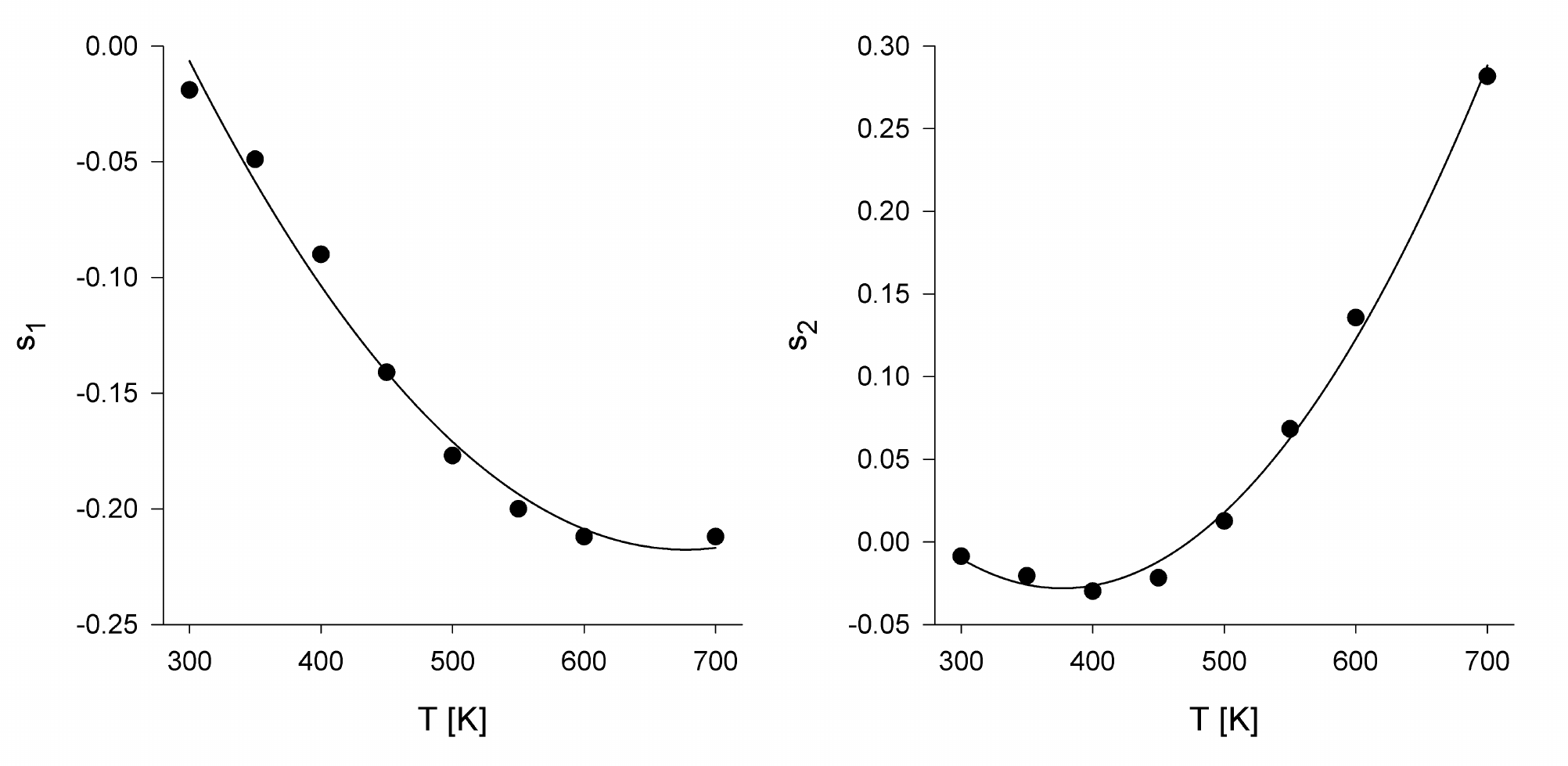}}
\caption{Temperature dependence of coefficients $s_i$.}
\label{fig:coef-nu}
\end{figure}

\begin{figure}[h!]
\centerline{\includegraphics[scale=0.7]{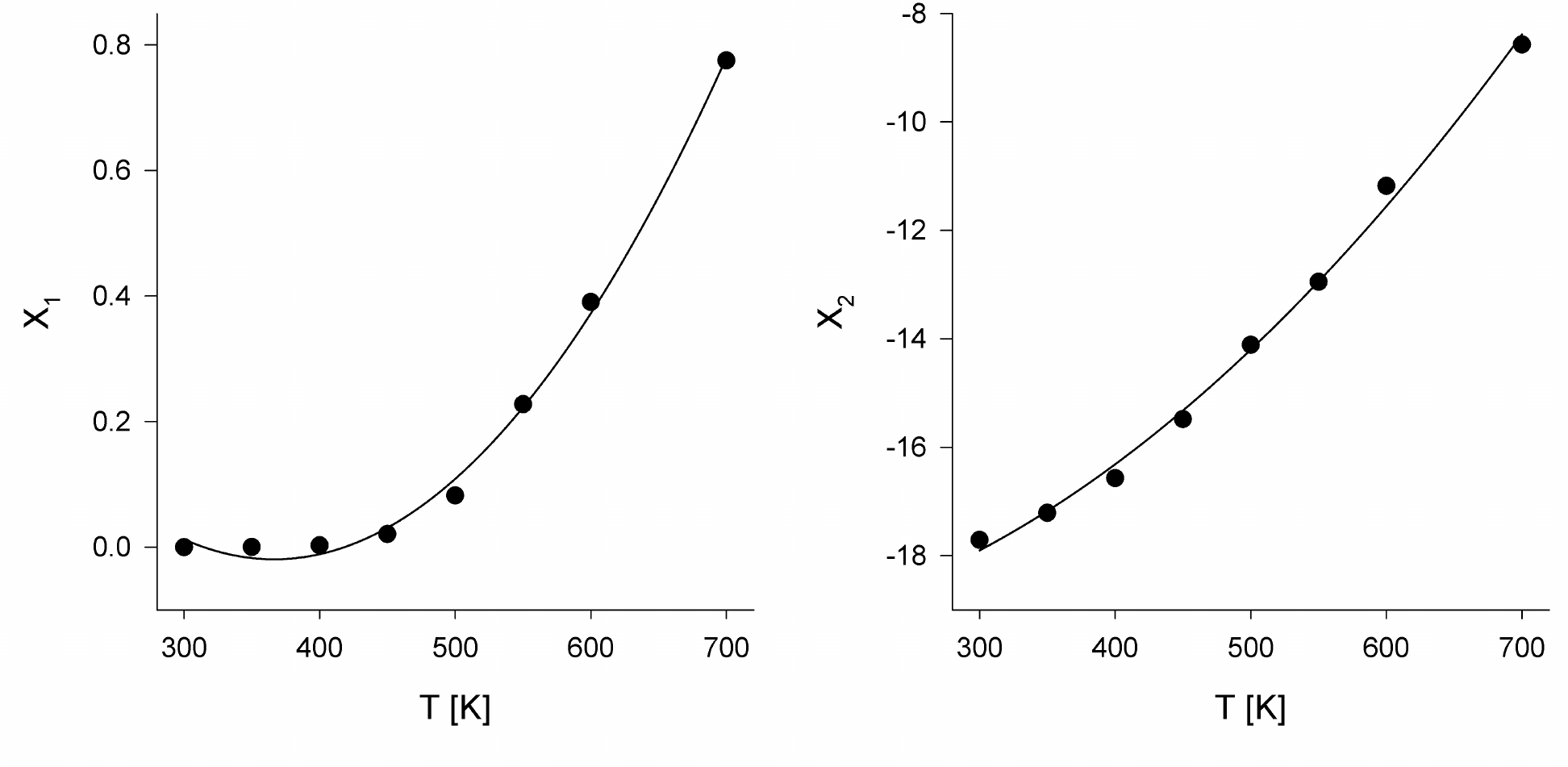}}
\caption{{Temperature dependence of coefficients $X_i$.}}
\label{fig:coef-x0}
\end{figure}

The above two functions, $\nu$ and $x_0$, contain two temperature dependent parameters which determine their curvature. The answer to the question, whether these parameters also exhibit a simple dependence on $T$ is provided in figures~\ref{fig:coef-nu} and~\ref{fig:coef-x0}.

The result is appealing: all these coefficients exhibit a simple quadratic dependence on $T$. In the end, we thus have at most 4$\times$3 constants available for fitting the experimental data.

\section{Conclusions}

In common applications of the SAFT methodology, the reference (primitive) model is defined by three parameters, the size of the hard sphere, and the range and strength of the H-bonding attraction. To gain a flexibility and improve the performance of the results, when fitting the experimental data, a temperature dependence of the HS diameter is introduced and the restrictions imposed by the TPT are lifted. These adjustments are rather arbitrary without physical considerations. As an attempt to incorporate/offer at least some physical insight into this amendment, we suggest that a tight link to the reference simple model should be broken by focussing on a general behavior of the derived properties, the parameter of non-saturation, $\gamma$ and $x_0$, instead of fitting directly the parameters of the reference model.
To summarize, a rather complicated computation of parameters $\gamma$ and $x_0$ for the given reference model may be replaced by general functions of the known temperature and density functional forms ensured by the physical nature.

\section*{Acknowledgement}

Support for this work was provided by the Czech Science Foundation (Grant No. 20-06825S).



	\ukrainianpart

\title{Термодинамічна теорія збурень та отримання рівняння стану}

\author[І. Незбеда]{І. Незбеда}

\address{Інститут фундаментальних основ хімічних процесів Чеської академії наук,
	16502 Прага 6, Чеська Республіка,\\
	Природничий факультет університету ім. Я.Є. Пуркіне,
	40096 Усті-над-Лабою, Чеська Республіка}

\makeukrtitle

\begin{abstract}
Альтернативний спосіб використання термодинамічної теорії збурень Вертгайма для отримання рівнянь стану асоціативних моделей флюїдів представлено та детально описано для води. Даний підхід використовує загальні характеристики параметра ненасичення, щоб уникнути розв'язування алгебраїчних рівнянь, та робить результати достатньо незалежними від деталей моделі простого референсного флюїду, що застосовується у теорії збурень.

\end{abstract}

\lastpage
\end{document}